\documentclass[10pt]{article}

\usepackage{graphicx}

\title{Simplified Variational Principles for non-Barotropic Magnetohydrodynamics}
\author{Asher Yahalom$^{a}$ \\
$^a$Ariel University, Ariel 40700, Israel\\
e-mail:  asya@ariel.ac.il; }

\begin{document}
\maketitle

\newcommand{\beq} {\begin{equation}}
\newcommand{\enq} {\end{equation}}
\newcommand{\ber} {\begin {eqnarray}}
\newcommand{\enr} {\end {eqnarray}}
\newcommand{\eq} {equation}
\newcommand{\eqn} {equation }
\newcommand{\eqs} {equations }
\newcommand{\ens} {equations}
\newcommand{\mn}  {{\mu \nu}}
\newcommand {\er}[1] {equation (\ref{#1}) }
\newcommand {\ern}[1] {equation (\ref{#1})}
\newcommand {\ers}[1] {equations (\ref{#1})}
\newcommand {\Er}[1] {Equation (\ref{#1}) }

\begin {abstract}

Variational principles for magnetohydrodynamics were introduced by
previous authors both in Lagrangian and Eulerian form. In this
paper we introduce simpler Eulerian variational principles from
which all the relevant
 equations of non-barotropic magnetohydrodynamics can be derived.
 The variational principle is given in terms of five independent functions
for non-stationary barotropic flows. This is less then the eight
variables which appear in the standard equations of barotropic
magnetohydrodynamics which are the magnetic field $\vec B$ the
velocity field $\vec v$, the entropy $s$ and the density $\rho$.

\vspace{0.3cm}

\vspace{0.3cm}
\noindent PACS number(s): 47.65.+a
\end {abstract}

\section {Introduction}

Variational principles for magnetohydrodynamics were introduced by
previous authors both in Lagrangian and Eulerian form. Sturrock
\cite{Sturrock} has discussed in his book a Lagrangian variational
formalism for magnetohydrodynamics. Vladimirov and Moffatt
\cite{Moffatt} in a series of papers have discussed an Eulerian
variational principle for incompressible magnetohydrodynamics.
However, their variational principle contained three more
functions in addition to the seven variables which appear in the
standard equations of incompressible magnetohydrodynamics which are the magnetic
field $\vec B$ the velocity field $\vec v$ and the pressure $P$.
Kats \cite{Kats} has generalized Moffatt's work for compressible
non barotropic flows but without reducing the number of functions
and the computational load. Moreover, Kats has shown that the
variables he suggested can be utilized to describe the motion of
arbitrary discontinuity surfaces \cite{Kats3,Kats4}. Sakurai
\cite{Sakurai} has introduced a two function Eulerian variational
principle for force-free magnetohydrodynamics and used it as a
basis of a numerical scheme, his method is discussed in a book by
Sturrock \cite{Sturrock}. A method of solving the equations for
those two variables was introduced by Yang, Sturrock \& Antiochos
\cite{Yang}. Yahalom \& Lynden-Bell \cite{YaLy} combined the Lagrangian of
Sturrock \cite{Sturrock} with the Lagrangian of Sakurai
\cite{Sakurai} to obtain an {\bf Eulerian} Lagrangian principle for barotropic magnetohydrodynamics
which will depend on only six functions. The variational
derivative of this Lagrangian produced all the equations
needed to describe barotropic magnetohydrodynamics without any
additional constraints. The equations obtained resembled the
equations of Frenkel, Levich \& Stilman \cite{FLS} (see also \cite{Zakharov}).
Yahalom \cite{Yah} have shown that for the barotropic case four functions will
suffice. Moreover, it was shown that the cuts of some of those functions \cite{Yah2}
are topological local conserved quantities.

Previous work was concerned only with
barotropic magnetohydrodynamics. Variational principles of non
barotropic magnetohydrodynamics can be found in the work of
Bekenstein \& Oron \cite{Bekenstien} in terms of 15 functions and
V.A. Kats \cite{Kats} in terms of 20 functions. The author of
this paper suspect that this number can be somewhat reduced.
Moreover, A. V. Kats  in a remarkable paper \cite{Kats2} (section
IV,E) has shown that there is a large symmetry group (gauge
freedom)  associated with the choice of those functions, this
implies that the number of degrees of freedom can be reduced.
Here we will show that only five functions will suffice to
describe non barotropic magnetohydrodynamics in the case
that we enforce a Sakurai \cite{Sakurai} representation for the magnetic field.

We anticipate applications of this study both to linear and
non-linear stability analysis of known non barotropic magnetohydrodynamic configurations \cite{VMI,AHH}
 and for designing efficient numerical schemes for integrating
the equations of fluid dynamics and magnetohydrodynamics
\cite{Yahalom,YahalomPinhasi, YahPinhasKop,OphirYahPinhasKop}.
Another possible application is connected to  obtaining new analytic solutions in terms
of the variational variables \cite{Yah3}.

The plan of this paper is as follows: First we introduce the
standard notations and equations of non-barotropic
magnetohydrodynamics. Next we introduce a generalization
of the barotropic variational principle suitable for the non-barotropic case.
Later we simplify the Eulerian variational principle and formulate it in terms of eight functions. Next we show
how three variational variables can be integrated algebraically thus reducing the variational principle
to five functions. We conclude by writing down the appropriate Hamiltonian for non-barotropic magnetohydrodynamics
and writing the equivalent Hamilton's equations for this case.

\section{Standard formulation of non-barotropic magnetohydrodynamics}

The standard set of \eqs solved for non-barotropic magnetohydrodynamics are given below:
\beq
\frac{\partial{\vec B}}{\partial t} = \vec \nabla \times (\vec v \times \vec B),
\label{Beq}
\enq
\beq
\vec \nabla \cdot \vec B =0,
\label{Bcon}
\enq
\beq
\frac{\partial{\rho}}{\partial t} + \vec \nabla \cdot (\rho \vec v ) = 0,
\label{masscon}
\enq
\beq
\rho \frac{d \vec v}{d t}=
\rho (\frac{\partial \vec v}{\partial t}+(\vec v \cdot \vec \nabla)\vec v)  = -\vec \nabla p (\rho,s) +
\frac{(\vec \nabla \times \vec B) \times \vec B}{4 \pi}.
\label{Euler}
\enq
\beq
 \frac{d s}{d t}=0.
\label{Ent}
\enq
The following notations are utilized: $\frac{\partial}{\partial t}$ is the temporal derivative,
$\frac{d}{d t}$ is the temporal material derivative and $\vec \nabla$ has its
standard meaning in vector calculus. $\vec B$ is the magnetic field vector, $\vec v$ is the
velocity field vector, $\rho$ is the fluid density and $s$ is the specific entropy. Finally $p (\rho,s)$ is the pressure which
depends on the density and entropy (the non-barotropic case). The justification for those \eqs
and the conditions under which they apply can be
found in standard books on magnetohydrodynamics (see for example \cite{Sturrock}). \Er{Beq}describes the
fact that the magnetic field lines are moving with the fluid elements ("frozen" magnetic field lines),
 \ern{Bcon} describes the fact that
the magnetic field is solenoidal, \ern{masscon} describes the conservation of mass and \ern{Euler}
is the Euler equation for a fluid in which both pressure
and Lorentz magnetic forces apply. The term:
\beq
\vec J =\frac{\vec \nabla \times \vec B}{4 \pi},
\label{J}
\enq
is the electric current density which is not connected to any mass flow.
\Er{Ent} describes the fact that heat is not created (zero viscosity, zero resistivity) in ideal non-barotropic magnetohydrodynamics
and is not conducted, thus only convection occurs.
The number of independent variables for which one needs to solve is eight
($\vec v,\vec B,\rho,s$) and the number of \eqs (\ref{Beq},\ref{masscon},\ref{Euler},\ref{Ent}) is also eight.
Notice that \ern{Bcon} is a condition on the initial $\vec B$ field and is satisfied automatically for
any other time due to \ern{Beq}.

\section{Variational principle of non-barotropic magnetohydrodynamics}

In the following section we will generalize the approach of \cite{YaLy} for the non-barotropic case.
Consider the action:
\ber A & \equiv & \int {\cal L} d^3 x dt,
\nonumber \\
{\cal L} & \equiv & {\cal L}_1 + {\cal L}_2,
\nonumber \\
{\cal L}_1 & \equiv & \rho (\frac{1}{2} \vec v^2 - \varepsilon (\rho,s)) +  \frac{\vec B^2}{8 \pi},
\nonumber \\
{\cal L}_2 & \equiv & \nu [\frac{\partial{\rho}}{\partial t} + \vec \nabla \cdot (\rho \vec v )]
- \rho \alpha \frac{d \chi}{dt} - \rho \beta \frac{d \eta}{dt} - \rho \sigma \frac{d s}{dt}
 - \frac{\vec B}{4 \pi} \cdot \vec \nabla \chi \times \vec \nabla \eta.
\label{Lagactionsimp}
\enr
Obviously $\nu,\alpha,\beta,\sigma$ are Lagrange multipliers which were inserted in such a
way that the variational principle will yield the following \ens:
\ber
& & \frac{\partial{\rho}}{\partial t} + \vec \nabla \cdot (\rho \vec v ) = 0,
\nonumber \\
& & \rho \frac{d \chi}{dt} = 0,
\nonumber \\
& & \rho \frac{d \eta}{dt} = 0.
\nonumber \\
& & \rho \frac{d s}{dt} = 0.
\label{lagmul}
\enr
It {\bf is not} assumed that $\nu,\alpha,\beta,\sigma$  are single valued.
Provided $\rho$ is not null those are just the continuity \ern{masscon}, entropy conservation
 and the conditions that Sakurai's functions are comoving.
Taking the variational derivative with respect to $\vec B$ we see that
\beq
\vec B = \hat {\vec B} \equiv \vec \nabla \chi \times \vec \nabla \eta.
\label{Bsakurai2}
\enq
Hence $\vec B$ is in Sakurai's form and satisfies \ern{Bcon}.
It can be easily shown that provided that $\vec B$ is in the form given in \ern{Bsakurai2},
and \ers{lagmul} are satisfied, then also \ern{Beq} is satisfied.

For the time being we have showed that all the equations of non-barotropic magnetohydrodynamics can be obtained
from the above variational principle except Euler's equations. We will now
show that Euler's equations can be derived from the above variational principle
as well. Let us take an arbitrary variational derivative of the above action with
respect to $\vec v$, this will result in:
\beq
\delta_{\vec v} A = \hspace{-0.2cm} \int dt \{ \int d^3 x dt \rho \delta \vec v \cdot
[\vec v - \vec \nabla \nu - \alpha \vec \nabla \chi - \beta \vec \nabla \eta - \sigma \vec \nabla s]
+ \oint d \vec S \cdot \delta \vec v \rho \nu+  \int d \vec \Sigma \cdot \delta \vec v \rho [\nu]\}.
\label{delActionv}
\enq
The integral $\oint d \vec S \cdot \delta \vec v \rho \nu$ vanishes in many physical scenarios.
In the case of astrophysical flows this integral will vanish since $\rho=0$ on the flow
boundary, in the case of a fluid contained
in a vessel no flux boundary conditions $\delta \vec v \cdot \hat n =0$ are induced
($\hat n$ is a unit vector normal to the boundary). The surface integral $\int d \vec \Sigma$
 on the cut of $\nu$ vanishes in the case that $\nu$ is single valued and $[\nu]=0$ .
In the case that $\nu$ is not single valued only a Kutta type velocity perturbation \cite{YahPinhasKop} in which
the velocity perturbation is parallel to the cut will cause the cut integral to vanish.

Provided that the surface integrals do vanish and that $\delta_{\vec v} A =0$ for an arbitrary
velocity perturbation we see that $\vec v$ must have the following form:
\beq
\vec v = \hat {\vec v} \equiv \vec \nabla \nu + \alpha \vec \nabla \chi + \beta \vec \nabla \eta + \sigma \vec \nabla s.
\label{vform}
\enq
The above equation is reminiscent of Clebsch representation in non magnetic fluids \cite{Clebsch1,Clebsch2}.
Let us now take the variational derivative with respect to the density $\rho$ we obtain:
\ber
\delta_{\rho} A & = & \int d^3 x dt \delta \rho
[\frac{1}{2} \vec v^2 - w  - \frac{\partial{\nu}}{\partial t} -  \vec v \cdot \vec \nabla \nu]
\nonumber \\
 & + & \int dt \oint d \vec S \cdot \vec v \delta \rho  \nu +
  \int dt \int d \vec \Sigma \cdot \vec v \delta \rho  [\nu] + \int d^3 x \nu \delta \rho |^{t_1}_{t_0}.
\label{delActionrho}
\enr
In which $ w= \frac{\partial (\varepsilon \rho)}{\partial \rho}$ is the specific enthalpy.
Hence provided that $\oint d \vec S \cdot \vec v \delta \rho  \nu$ vanishes on the boundary of the domain
and $ \int d \vec \Sigma \cdot \vec v \delta \rho  [\nu]$ vanishes on the cut of $\nu$
in the case that $\nu$ is not single valued\footnote{Which entails either a Kutta type
condition for the velocity or a vanishing density perturbation on the cut.}
and in initial and final times the following \eqn must be satisfied:
\beq
\frac{d \nu}{d t} = \frac{1}{2} \vec v^2 - w, \qquad
\label{nueq}
\enq
Finally we have to calculate the variation with respect to both $\chi$ and $\eta$
this will lead us to the following results:
\ber
\delta_{\chi} A \hspace{-0.4cm} & = & \hspace{-0.4cm} \int d^3 x dt \delta \chi
[\frac{\partial{(\rho \alpha)}}{\partial t} +  \vec \nabla \cdot (\rho \alpha \vec v)-
\vec \nabla \eta \cdot \vec J]
+ \int dt \oint d \vec S \cdot [\frac{\vec B}{4 \pi} \times \vec \nabla \eta - \vec v \rho \alpha]\delta \chi
 \nonumber \\
 & + & \int dt \int d \vec \Sigma \cdot [\frac{\vec B}{4 \pi} \times \vec \nabla \eta - \vec v \rho \alpha][\delta \chi]
 - \int d^3 x \rho \alpha \delta \chi |^{t_1}_{t_0},
\label{delActionchi}
\enr
\ber
\delta_{\eta} A \hspace{-0.4cm} & = & \hspace{-0.4cm} \int d^3 x dt \delta \eta
[\frac{\partial{(\rho \beta)}}{\partial t} +  \vec \nabla \cdot (\rho \beta \vec v)+
\vec \nabla \chi \cdot \vec J]
+ \int dt \oint d \vec S \cdot [\vec \nabla \chi \times \frac{\vec B}{4 \pi} - \vec v \rho \beta]\delta \eta
\nonumber \\
 & + &  \int dt \int d \vec \Sigma \cdot [\vec \nabla \chi \times \frac{\vec B}{4 \pi} - \vec v \rho \beta][\delta \eta]
 - \int d^3 x \rho \beta \delta \eta |^{t_1}_{t_0}.
\label{delActioneta}
\enr
Provided that the correct temporal and boundary conditions are met with
respect to the variations $\delta \chi$ and $\delta \eta$ on the domain boundary and
on the cuts in the case that some (or all) of the relevant functions are non single valued.
we obtain the following set of equations:
\beq
\frac{d \alpha}{dt} = \frac{\vec \nabla \eta \cdot \vec J}{\rho}, \qquad
\frac{d \beta}{dt} = -\frac{\vec \nabla \chi \cdot \vec J}{\rho},
\label{albetaeq}
\enq
in which the continuity \ern{masscon} was taken into account. By correct temporal conditions we
mean that both $\delta \eta$ and $\delta \chi$ vanish at initial and final times. As for boundary
conditions which are sufficient to make the boundary term vanish on can consider the case that
the boundary is at infinity and both $\vec B$ and $\rho$ vanish. Another possibility is that the boundary is
impermeable and perfectly conducting. A sufficient condition for the integral over the "cuts" to vanish
is to use variations $\delta \eta$ and $\delta \chi$ which are single valued. It can be shown that
$\chi$ can always be taken to be single valued, hence taking $\delta \chi$ to be single valued is no
restriction at all. In some topologies $\eta$ is not single valued and in those cases a single valued
restriction on $\delta \eta$ is sufficient to make the cut term null.

Finally we take a variational derivative with respect to the entropy $s$:
\ber
\delta_{s} A \hspace{-0.4cm} & = & \hspace{-0.4cm} \int d^3 x dt \delta s
[\frac{\partial{(\rho \sigma)}}{\partial t} +  \vec \nabla \cdot (\rho \sigma \vec v)- \rho T]
+ \int dt \oint d \vec S \cdot \rho \sigma \vec v  \delta s
 \nonumber \\
 & - &  \int d^3 x \rho \sigma \delta s |^{t_1}_{t_0},
\label{delActions}
\enr
in which the temperature is $T=\frac{\partial \varepsilon}{\partial s}$. We notice that according
to \ern{vform} $\sigma$ is single valued and hence no cuts are needed. Taking into account the continuity
\ern{masscon} we obtain for locations in which the density $\rho$ is not null the result:
\beq
\frac{d \sigma}{dt} =T,
\label{sigmaeq}
\enq
provided that $\delta_{s} A$ vanished for an arbitrary $\delta s$.

\section{Euler's equations}

We shall now show that a velocity field given by \ern{vform}, such that the
\eqs for $\alpha, \beta, \chi, \eta, \nu, \sigma, s$ satisfy the corresponding equations
(\ref{lagmul},\ref{nueq},\ref{albetaeq},\ref{sigmaeq}) must satisfy Euler's equations.
Let us calculate the material derivative of $\vec v$:
\beq
\frac{d\vec v}{dt} = \frac{d\vec \nabla \nu}{dt}  + \frac{d\alpha}{dt} \vec \nabla \chi +
 \alpha \frac{d\vec \nabla \chi}{dt}  +
\frac{d\beta}{dt} \vec \nabla \eta + \beta \frac{d\vec \nabla \eta}{dt}+\frac{d\sigma}{dt} \vec \nabla s +
\sigma \frac{d\vec \nabla s}{dt}.
\label{dvform}
\enq
It can be easily shown that:
\ber
\frac{d\vec \nabla \nu}{dt} & = & \vec \nabla \frac{d \nu}{dt}- \vec \nabla v_k \frac{\partial \nu}{\partial x_k}
 = \vec \nabla (\frac{1}{2} \vec v^2 - w)- \vec \nabla v_k \frac{\partial \nu}{\partial x_k},
 \nonumber \\
 \frac{d\vec \nabla \eta}{dt} & = & \vec \nabla \frac{d \eta}{dt}- \vec \nabla v_k \frac{\partial \eta}{\partial x_k}
 = - \vec \nabla v_k \frac{\partial \eta}{\partial x_k},
 \nonumber \\
 \frac{d\vec \nabla \chi}{dt} & = & \vec \nabla \frac{d \chi}{dt}- \vec \nabla v_k \frac{\partial \chi}{\partial x_k}
 = - \vec \nabla v_k \frac{\partial \chi}{\partial x_k},
  \nonumber \\
 \frac{d\vec \nabla s}{dt} & = & \vec \nabla \frac{d s}{dt}- \vec \nabla v_k \frac{\partial s}{\partial x_k}
 = - \vec \nabla v_k \frac{\partial s}{\partial x_k}.
 \label{dnabla}
\enr
In which $x_k$ is a Cartesian coordinate and a summation convention is assumed. Inserting the result from equations (\ref{dnabla},\ref{lagmul})
into \ern{dvform} yields:
\ber
\frac{d\vec v}{dt} &=& - \vec \nabla v_k (\frac{\partial \nu}{\partial x_k} + \alpha \frac{\partial \chi}{\partial x_k} +
\beta \frac{\partial \eta}{\partial x_k} + \sigma \frac{\partial s}{\partial x_k}) + \vec \nabla (\frac{1}{2} \vec v^2 - w)+ T \vec \nabla s
 \nonumber \\
&+& \frac{1}{\rho} ((\vec \nabla \eta \cdot \vec J)\vec \nabla \chi - (\vec \nabla \chi \cdot \vec J)\vec \nabla \eta)
 \nonumber \\
&=& - \vec \nabla v_k v_k + \vec \nabla (\frac{1}{2} \vec v^2 - w) + T \vec \nabla s
 + \frac{1}{\rho} \vec J \times (\vec \nabla \chi \times  \vec \nabla \eta)
 \nonumber \\
&=& - \frac{\vec \nabla p}{\rho} + \frac{1}{\rho} \vec J \times \vec B.
\label{dvform2}
\enr
In which we have used both \ern{vform} and \ern{Bsakurai2} in the above derivation. This of course
proves that the non-barotropic Euler equations can be derived from the action given in \er{Lagactionsimp} and hence
all the equations of non-barotropic magnetohydrodynamics can be derived from the above action
without restricting the variations in any way except on the relevant boundaries and cuts.

\section{Simplified action}

The reader of this paper might argue here that the paper is misleading. The author has declared
that he is going to present a simplified action for non-barotropic magnetohydrodynamics instead he
 added six more functions $\alpha,\beta,\chi,\-\eta,\nu,\sigma$ to the standard set $\vec B,\vec v,\rho,s$.
In the following I will show that this is not so and the action given in \ern{Lagactionsimp} in
a form suitable for a pedagogic presentation can indeed be simplified. It is easy to show
that the Lagrangian density appearing in \ern{Lagactionsimp} can be written in the form:
\ber
{\cal L} & = & -\rho [\frac{\partial{\nu}}{\partial t} + \alpha \frac{\partial{\chi}}{\partial t}
+ \beta \frac{\partial{\eta}}{\partial t}+ \sigma \frac{\partial{s}}{\partial t}+\varepsilon (\rho,s)] +
\frac{1}{2}\rho [(\vec v-\hat{\vec v})^2-(\hat{\vec v})^2]
\nonumber \\
& + &   \frac{1}{8 \pi} [(\vec B-\hat{\vec B})^2-(\hat{\vec B})^2]+
\frac{\partial{(\nu \rho)}}{\partial t} + \vec \nabla \cdot (\nu \rho \vec v ).
\label{Lagactionsimp4}
\enr
In which $\hat{\vec v}$ is a shorthand notation for $\vec \nabla \nu + \alpha \vec \nabla \chi +
 \beta \vec \nabla \eta +  \sigma \vec \nabla s $ (see \ern{vform}) and $\hat{\vec B}$ is a shorthand notation for
 $\vec \nabla \chi \times \vec \nabla \eta$ (see \ern{Bsakurai2}). Thus ${\cal L}$ has four contributions:
\ber
  {\cal L}  &  = &  \hat {\cal L} + {\cal L}_{\vec v}+ {\cal L}_{\vec B}+{\cal L}_{boundary},
\nonumber \\
\hat {\cal L}   \hspace{1.2 cm} &\hspace{-2.5 cm} \equiv &  \hspace{-1.5 cm} -\rho \left[\frac{\partial{\nu}}{\partial t} + \alpha \frac{\partial{\chi}}{\partial t}
+ \beta \frac{\partial{\eta}}{\partial t}+ \sigma \frac{\partial{s}}{\partial t}+\varepsilon (\rho,s)+
\frac{1}{2} (\vec \nabla \nu + \alpha \vec \nabla \chi +  \beta \vec \nabla \eta +  \sigma \vec \nabla s )^2 \right]
\nonumber \\
&-&\frac{1}{8 \pi}(\vec \nabla \chi \times \vec \nabla \eta)^2
\nonumber \\
{\cal L}_{\vec v} &\equiv & \frac{1}{2}\rho (\vec v-\hat{\vec v})^2,
\nonumber \\
{\cal L}_{\vec B} &\equiv & \frac{1}{8 \pi} (\vec B-\hat{\vec B})^2,
\nonumber \\
{\cal L}_{boundary} &\equiv & \frac{\partial{(\nu \rho)}}{\partial t} + \vec \nabla \cdot (\nu \rho \vec v ).
\label{Lagactionsimp5}
\enr
The only term containing $\vec v$ is\footnote{${\cal L}_{boundary}$ also depends on
$\vec v$ but being a boundary term is space and time it does not contribute to the derived equations}
 ${\cal L}_{\vec v}$, it can easily be seen that
this term will lead, after we nullify the variational derivative with respect to $\vec v$,
to \ern{vform} but will otherwise
have no contribution to other variational derivatives. Similarly the only term containing $\vec B$
is ${\cal L}_{\vec B}$ and it can easily be seen that
this term will lead, after we nullify the variational derivative, to \ern{Bsakurai2} but will
have no contribution to other variational derivatives. Also notice that the term ${\cal L}_{boundary}$
contains only complete partial derivatives and thus can not contribute to the equations although
it can change the boundary conditions. Hence we see that \ers{lagmul}, \ern{nueq}, \ers{albetaeq} and \er{sigmaeq}
can be derived using the Lagrangian density:
\ber
& & \hat {\cal L}[\alpha,\beta,\chi,\eta,\nu,\rho,\sigma,s] = -\rho [\frac{\partial{\nu}}{\partial t} + \alpha \frac{\partial{\chi}}{\partial t}
+ \beta \frac{\partial{\eta}}{\partial t}+ \sigma \frac{\partial{s}}{\partial t}
\nonumber \\
& & +\  \varepsilon (\rho,s) + \frac{1}{2} (\vec \nabla \nu + \alpha \vec \nabla \chi +  \beta \vec \nabla \eta +  \sigma \vec \nabla s )^2 ]
-\frac{1}{8 \pi}(\vec \nabla \chi \times \vec \nabla \eta)^2
\label{Lagactionsimp6}
\enr
in which $\hat{\vec v}$ replaces $\vec v$ and $\hat{\vec B}$ replaces $\vec B$ in the relevant equations.
Furthermore, after integrating the eight \eqs
(\ref{lagmul},\ref{nueq},\ref{albetaeq},\ref{sigmaeq}) we can insert the potentials $\alpha,\beta,\chi,\eta,\nu,\sigma,s$
into \ers{vform} and (\ref{Bsakurai2}) to obtain the physical quantities $\vec v$ and $\vec B$.
Hence, the general non-barotropic magnetohydrodynamic problem is reduced from eight equations
(\ref{Beq},\ref{masscon},\ref{Euler},\ref{Ent}) and the additional constraint (\ref{Bcon})
to a problem of eight first order (in the temporal derivative) unconstrained equations.
Moreover, the entire set of equations can be derived from the Lagrangian density $\hat {\cal L}$.

\section{Further Simplification}

\subsection{Elimination of Variables}

 Let us now look at the three last three equations of (\ref{lagmul}). Those describe three comoving quantities
 which can be written in terms of the generalized Clebsch form given in \ern{vform} as follows:
\ber
& &  \frac{\partial \chi}{\partial t} + (\vec \nabla \nu + \alpha \vec \nabla \chi + \beta \vec \nabla \eta + \sigma \vec \nabla s)
\cdot \vec \nabla \chi = 0
\nonumber \\
& & \frac{\partial \eta}{\partial t} + (\vec \nabla \nu + \alpha \vec \nabla \chi + \beta \vec \nabla \eta + \sigma \vec \nabla s)
\cdot \vec \nabla \eta = 0
\nonumber \\
& & \frac{\partial s}{\partial t} + (\vec \nabla \nu + \alpha \vec \nabla \chi + \beta \vec \nabla \eta + \sigma \vec \nabla s)
\cdot \vec \nabla s = 0
\label{lagmul4}
\enr
Those are algebraic equations for $\alpha, \beta, \sigma$. Which can be solved such that $\alpha, \beta, \sigma$ can be written
as functionals of $\chi,\eta,\nu,s$, resulting eventually in the description of non-barotropic magnetohydrodynamics
in terms of five functions: $\nu,\rho,\chi,\eta,s$.
Let us introduce the notation:
\beq
\alpha_i \equiv (\alpha, \beta, \sigma), \quad \chi_i\equiv (\chi,\eta,s), \quad
 k_i \equiv -\frac{\partial \chi_i}{\partial t} - \vec \nabla \nu \cdot \vec \nabla \chi_i, \qquad
i\in(1,2,3)
 \label{ali}
\enq
In terms of the above notation \ern{lagmul4} takes the form:
\beq
k_i =\alpha_j \vec \nabla \chi_i \cdot \vec \nabla \chi_j, \qquad j\in(1,2,3)
\label{kieq}
\enq
in which the Einstein summation convention is assumed. Let us define the matrix:
\beq
A_{ij} \equiv  \vec \nabla \chi_i \cdot \vec \nabla \chi_j
\label{Adef}
\enq
obviously this matrix is symmetric since $A_{ij}=A_{ji}$. Hence \er{kieq} takes the form:
\beq
k_i = A_{ij} \alpha_j, \qquad j\in(1,2,3)
\label{kieq2}
\enq
 Provided that the matrix $A_{ij}$ is not singular it has an inverse $A^{-1}_{ij}$ which can be written as:
\beq
A^{-1}_{ij}=\left|A\right|^{-1} \left(
\begin{array}{ccc}
 A_{22} A_{33}-A_{23}^2 & A_{13} A_{23}-A_{12} A_{33} & A_{12} A_{23}-A_{13} A_{22} \\
 A_{13} A_{23}-A_{12} A_{33} & A_{11} A_{33}-A_{13}^2 & A_{12} A_{13}-A_{11} A_{23} \\
 A_{12} A_{23}-A_{13} A_{22} & A_{12} A_{13}-A_{11} A_{23} & A_{11} A_{22}-A_{12}^2
\end{array}
\right)
\label{invAdef}
\enq
In which the determinant $\left|A\right|$ is given by the following equation:
\beq
\left|A\right|=
A_{11} A_{22} A_{33}-A_{11} A_{23}^2-A_{22} A_{13}^2 -A_{33} A_{12}^2 +2 A_{12} A_{13} A_{23}
\label{Adet}
\enq
In terms of the above equations the $\alpha_i$'s can be calculated as functionals of $\chi_i,\nu$ as
follows:
\beq
\alpha_i [\chi_i,\nu]= A^{-1}_{ij} k_j.
\label{aleq}
\enq
The velocity \ern{vform} can now be written as:
\beq
\vec v = \vec \nabla \nu + \alpha_i \vec \nabla \chi_i=  \vec \nabla \nu + A^{-1}_{ij} k_j \vec \nabla \chi_i
=\vec \nabla \nu - A^{-1}_{ij}\vec \nabla \chi_i (\frac{\partial \chi_j}{\partial t} + \vec \nabla \nu \cdot \vec \nabla \chi_j).
\label{vform2}
\enq
Provided that the $\chi_i$ is a coordinate basis in three dimensions, we may write:
\beq
\vec \nabla \nu= \vec \nabla \chi_n \frac{\partial \nu}{\partial \chi_n}, \qquad n\in(1,2,3).
\label{nudecom}
\enq
Inserting \ern{nudecom} into \ern{vform2} we obtain:
\ber
\vec v &=& - A^{-1}_{ij}\vec \nabla \chi_i \frac{\partial \chi_j}{\partial t}+
\vec \nabla \nu - A^{-1}_{ij}\vec \nabla \chi_i  \frac{\partial \nu}{\partial \chi_n} \vec \nabla \chi_n \cdot \vec \nabla \chi_j
\nonumber \\
&=& - A^{-1}_{ij}\vec \nabla \chi_i \frac{\partial \chi_j}{\partial t}+
\vec \nabla \nu - A^{-1}_{ij} A_{jn} \vec \nabla \chi_i  \frac{\partial \nu}{\partial \chi_n}
\nonumber \\
&=& - A^{-1}_{ij}\vec \nabla \chi_i \frac{\partial \chi_j}{\partial t}+
\vec \nabla \nu - \delta_{in} \vec \nabla \chi_i  \frac{\partial \nu}{\partial \chi_n}
\nonumber \\
&=& - A^{-1}_{ij}\vec \nabla \chi_i \frac{\partial \chi_j}{\partial t}+
\vec \nabla \nu -  \vec \nabla \chi_n  \frac{\partial \nu}{\partial \chi_n}
\nonumber \\
&=& - A^{-1}_{ij}\vec \nabla \chi_i \frac{\partial \chi_j}{\partial t}
\label{vform3}
\enr
in the above $\delta_{in}$ is a Kronecker delta. Thus the velocity $\vec v [\chi_i]$ is a functional of
$\chi_i$ only and is independent of $\nu$.

\subsection{Lagrangian Density and Variational Analysis}

Let us now rewrite the Lagrangian density $\hat {\cal L}[\chi_i,\nu,\rho]$ given in
\ern{Lagactionsimp6} in terms of the new variables:
\beq
 \hat {\cal L}[\chi_i,\nu,\rho] = -\rho [\frac{\partial{\nu}}{\partial t} + \alpha_k [\chi_i,nu] \frac{\partial{\chi_k}}{\partial t}
 +\  \varepsilon (\rho,\chi_3) + \frac{1}{2} \vec v [\chi_i]^2 ]
-\frac{1}{8 \pi}(\vec \nabla \chi_1 \times \vec \nabla \chi_2)^2
\label{Lagactionsimp7}
\enq
Let us calculate the variational derivative of $\hat {\cal L}[\chi_i,\nu,\rho]$ with respect to $\chi_i$ this will result in:
\beq
 \delta_{\chi_i}\hat {\cal L} = -\rho [  \delta_{\chi_i} \alpha_k  \frac{\partial{\chi_k}}{\partial t} +
\alpha_{\underline{i}}  \frac{\partial \delta \chi_{\underline{i}}}{\partial t}
 +\  \delta_{\chi_i} \varepsilon (\rho,\chi_3) +  \delta_{\chi_i}\vec v \cdot \vec v ]
-\frac{ \vec B} {4 \pi} \cdot  \delta_{\chi_i} (\vec \nabla \chi_1 \times \vec \nabla \chi_2)
\label{delchiLag}
\enq
in which the summation convention is not applied if the index is underlined.
However, due to \ern{vform2} we may write:
\beq
  \delta_{\chi_i}\vec v= \delta_{\chi_i} \alpha_k \vec \nabla \chi_k +   \alpha_{\underline{i}} \vec \nabla \delta \chi_{\underline{i}}.
\label{delchiv}
\enq
Inserting \ern{delchiv} into \ern{delchiLag} and rearranging the terms we obtain:
\ber
 \delta_{\chi_i}\hat {\cal L} &=& -\rho [  \delta_{\chi_i} \alpha_k  (\frac{\partial{\chi_k}}{\partial t}
+ \vec v \cdot \vec \nabla \chi_k )+
\alpha_{\underline{i}}  (\frac{\partial \delta \chi_{\underline{i}}}{\partial t}+ \vec v \cdot \vec \nabla \delta \chi_{\underline{i}})
 +\  \delta_{\chi_i} \varepsilon (\rho,\chi_3) ]
 \nonumber \\
&-& \frac{ \vec B} {4 \pi} \cdot  \delta_{\chi_i} (\vec \nabla \chi_1 \times \vec \nabla \chi_2).
\label{delchiLag2}
\enr
Now by construction $\vec v$ satisfies \ern{lagmul4} and hence $\frac{\partial{\chi_k}}{\partial t}
+ \vec v \cdot \vec \nabla \chi_k  = 0$, this leads to:
\beq
 \delta_{\chi_i}\hat {\cal L} = -\rho \left[  \alpha_{\underline{i}} \frac{d \delta \chi_{\underline{i}}}{d t}
  + \delta_{\chi_i} \varepsilon (\rho,\chi_3) \right] - \frac{ \vec B} {4 \pi} \cdot  \delta_{\chi_i} (\vec \nabla \chi_1 \times \vec \nabla \chi_2).
\label{delchiLag3}
\enq
From now on the derivation proceeds as in \eqs (\ref{delActionchi},\ref{delActioneta},\ref{delActions}) resulting in \eqs
(\ref{albetaeq},\ref{sigmaeq}) and will not be repeated. The difference is that now $\alpha, \beta$ and $\sigma$ are
 not independent quantities, rather they depend through \ern{aleq}
on the derivatives of $\chi_i,\nu$. Thus, \eqs (\ref{delActionchi},\ref{delActioneta},\ref{delActions})
 are not first order equations in time but are second order equations. Now let us calculate the variational derivative
 with respect to $\nu$ this will result in the expression:
\beq
 \delta_{\nu} \hat {\cal L} = -\rho [ \frac{\partial{\delta \nu}}{\partial t} + \delta_{\nu} \alpha_n  \frac{\partial{\chi_n}}{\partial t}]
\label{delnuLag}
\enq
However, $\delta_{\nu} \alpha_k$ can be calculated from \ern{aleq}:
\beq
\delta_{\nu} \alpha_n = A^{-1}_{nj} \delta_{\nu} k_j = - A^{-1}_{nj} \vec \nabla \delta \nu \cdot \vec \nabla \chi_j
\label{delnualeq}
\enq
Inserting the above equation into \ern{delnuLag}:
\beq
 \delta_{\nu} \hat {\cal L} = -\rho [ \frac{\partial{\delta \nu}}{\partial t} - A^{-1}_{nj}  \vec \nabla \chi_j
  \frac{\partial{\chi_n}}{\partial t} \cdot \vec \nabla \delta \nu ] =
  -\rho [ \frac{\partial{\delta \nu}}{\partial t} +  \vec v \cdot \vec \nabla \delta \nu ]=
   -\rho \frac{d{\delta \nu}}{d t}
\label{delnuLag2}
\enq
The above equation can be put to the form:
\beq
 \delta_{\nu} \hat {\cal L} = \delta \nu [\frac{\partial{\rho}}{\partial t} + \vec \nabla \cdot (\rho \vec v )]
-\frac{\partial{(\rho \delta \nu)}}{\partial t}- \vec \nabla \cdot (\rho \vec v \delta \nu )
\label{delnuLag3}
\enq
This obviously leads to the continuity \ern{masscon} and some boundary terms in space and time. The variational
derivative with respect to $\rho$ is trivial and the analysis is identical to the one in \ern{delActionrho} leading
to \ern{nueq}. To conclude this subsection let us summarize the equations of non-barotropic magnetohydrodynamics:
\ber
\frac{d \nu}{d t} &=& \frac{1}{2} \vec v^2 - w,
\nonumber \\
\frac{\partial{\rho}}{\partial t} &+& \vec \nabla \cdot (\rho \vec v ) = 0
\nonumber \\
\frac{d \sigma}{dt} &=& T,
\nonumber \\
\frac{d \alpha}{dt} &=& \frac{\vec \nabla \eta \cdot \vec J}{\rho},
\nonumber \\
\frac{d \beta}{dt} &=& -\frac{\vec \nabla \chi \cdot \vec J}{\rho},
\label{equa}
\enr
in which $\alpha,\beta,\sigma,\vec v$ are functionals of $\chi,\eta,s,\nu$ as described above.
It is easy to show as in \ern{dvform2} that those variational equations are equivalent to the physical equations.

\subsection{Lagrangian and Hamiltonian Densities}

Let us put the Lagrangian density of \er{Lagactionsimp7}in a slightly more explicit form. First us look at the
term $\vec v^2$:
\beq
\vec v^2 =
 A^{-1}_{ij}\vec \nabla \chi_i \frac{\partial \chi_j}{\partial t} A^{-1}_{mn}\vec \nabla \chi_m \frac{\partial \chi_n}{\partial t}=
A^{-1}_{ij} A^{-1}_{mn} A_{im} \frac{\partial \chi_j}{\partial t}  \frac{\partial \chi_n}{\partial t}=
A^{-1}_{jn} \frac{\partial \chi_j}{\partial t}  \frac{\partial \chi_n}{\partial t}
\label{vsq}
\enq
in the above we use \ern{vform3} and \ern{Adef}. Next let us look at the expression:
\ber
\alpha_k [\chi_i,\nu] \frac{\partial{\chi_k}}{\partial t}& =& A^{-1}_{kj} k_j \frac{\partial{\chi_k}}{\partial t}
 =-(\frac{\partial \chi_j}{\partial t} + \vec \nabla \nu \cdot \vec \nabla \chi_j)A^{-1}_{kj} \frac{\partial{\chi_k}}{\partial t}
 \nonumber \\
 &=& -A^{-1}_{jk} \frac{\partial \chi_j}{\partial t}  \frac{\partial \chi_k}{\partial t}
 -\vec \nabla \nu \cdot \vec \nabla \chi_j A^{-1}_{kj} \frac{\partial{\chi_k}}{\partial t}
  \nonumber \\
 &=& -A^{-1}_{jk} \frac{\partial \chi_j}{\partial t}  \frac{\partial \chi_k}{\partial t}
 - \frac{\partial \nu}{\partial \chi_m} \vec \nabla \chi_m \cdot \vec \nabla \chi_j A^{-1}_{kj} \frac{\partial{\chi_k}}{\partial t}
 \nonumber \\
 &=& -A^{-1}_{jk} \frac{\partial \chi_j}{\partial t}  \frac{\partial \chi_k}{\partial t}
 - \frac{\partial \nu}{\partial \chi_m} A_{mj} A^{-1}_{kj} \frac{\partial{\chi_k}}{\partial t}
 \nonumber \\
 &=& -A^{-1}_{jk} \frac{\partial \chi_j}{\partial t}  \frac{\partial \chi_k}{\partial t}
 - \frac{\partial \nu}{\partial \chi_m} \delta_{km} \frac{\partial{\chi_k}}{\partial t}
 \nonumber \\
 &=& -A^{-1}_{jk} \frac{\partial \chi_j}{\partial t}  \frac{\partial \chi_k}{\partial t}
 - \frac{\partial \nu}{\partial \chi_m} \frac{\partial{\chi_m}}{\partial t}
\label{alpterm}
\enr
Inserting \er{vsq} and \er{alpterm} into \er{Lagactionsimp7} leads to a Lagrangian density of a  more
standard quadratic form:
\beq
 \hat {\cal L}[\chi_i,\nu,\rho] = \rho [\frac{1}{2} A^{-1}_{jn} \frac{\partial \chi_j}{\partial t}  \frac{\partial \chi_n}{\partial t}
+\frac{\partial \nu}{\partial \chi_m} \frac{\partial{\chi_m}}{\partial t}-
 \frac{\partial{\nu}}{\partial t} -\  \varepsilon (\rho,\chi_3)]
-\frac{1}{8 \pi}(\vec \nabla \chi_1 \times \vec \nabla \chi_2)^2.
\label{Lagactionsimp8}
\enq
In which $A^{-1}_{jn}$ plays the rule of a "metric". The Lagrangian is thus composed of a kinetic terms which is quadratic in the temporal
derivatives, a "gyroscopic" terms which is linear in the temporal derivative and a potential term which is independent of the temporal derivative.

In order to obtain a Hamiltonian density of a more convenient form we will add a temporal derivative to the Lagrangian density $\hat {\cal L}$,
this will not change the dynamical equations as is well known, hence we can write:
\ber
 & & \hspace{-1cm} \tilde {\cal L}[\chi_i,\nu,\rho] = \hat {\cal L}+ \frac{\partial (\rho \nu)}{\partial t} =
\nonumber \\
& & \hspace{-1cm} \nu \frac{\partial \rho}{\partial t}
 +\rho [\frac{1}{2} A^{-1}_{jn} \frac{\partial \chi_j}{\partial t}  \frac{\partial \chi_n}{\partial t}
+\frac{\partial \nu}{\partial \chi_m} \frac{\partial{\chi_m}}{\partial t} - \varepsilon (\rho,\chi_3)]
-\frac{1}{8 \pi}(\vec \nabla \chi_1 \times \vec \nabla \chi_2)^2.
\label{Lagactionsimp9}
\enr
From this we obtain the following canonical momenta:
\beq
p_{\rho} = \frac{\partial \tilde {\cal L}}{\partial \dot{\rho}} = \nu, \qquad  \dot{\rho} \equiv \frac{\partial \rho}{\partial t}.
\label{prho}
\enq
And:
\beq
p_{\chi_k} = \frac{\partial \tilde {\cal L}}{\partial \dot{\chi}_k} =\rho[A^{-1}_{kj} \frac{\partial \chi_j}{\partial t}+
\frac{\partial \nu}{\partial \chi_k}]=-\rho \alpha_k,
\qquad  \dot{\chi}_k \equiv \frac{\partial \chi_k}{\partial t}.
\label{pchik}
\enq
The Hamiltonian density can be now calculated as follows:
\ber
{\cal H} [\chi_i,\rho,p_{\chi_k},p_{\rho}] &=& p_{\chi_k} \dot{\chi}_k + p_{\rho} \dot{\rho} - \tilde {\cal L}
\nonumber \\
&=&  \rho [\frac{1}{2} A^{-1}_{jn} \frac{\partial \chi_j}{\partial t}  \frac{\partial \chi_n}{\partial t} + \varepsilon (\rho,\chi_3)]
+\frac{1}{8 \pi}(\vec \nabla \chi_1 \times \vec \nabla \chi_2)^2.
\nonumber \\
& =& \frac{1}{2} \rho^{-1} A_{kl} p_{\chi_k} p_{\chi_l} - p_{\chi_k} \vec \nabla \chi_k \cdot \vec \nabla p_{\rho}+
\frac{1}{2} \rho (\vec \nabla p_{\rho})^2
\nonumber \\
&+& \rho \varepsilon (\rho,\chi_3) +\frac{1}{8 \pi}(\vec \nabla \chi_1 \times \vec \nabla \chi_2)^2.
\label{Hdensity}
\enr
Hence by virtue of \ern{vsq} the Hamiltonian density is equal to the energy density of the flow:
\beq
{\cal H}= \frac{1}{2} \rho \vec v^2 + \rho \varepsilon +\frac{\vec B^2}{8 \pi}
\label{Energydensity}
\enq
Hamilton equations are now:
\ber
& & \frac{\delta \cal H}{\delta p_{\rho}}= \dot{\rho}
\label{rhodot} \\
&-& \frac{\delta \cal H}{\delta \rho}= \dot{p}_{\rho}
\label{prhodot} \\
& & \frac{\delta \cal H}{\delta p_{\chi_k}}= \dot{\chi}_k
\label{chidot} \\
&-& \frac{\delta \cal H}{\delta\chi_k}= \dot{p}_{\chi_k}.
\label{pchidot} \\
\enr
\Er{rhodot} is equivalent to the continuity \ern{masscon}. \Er{prhodot} is equivalent to the Bernoulli type \ern{nueq}.
\Er{chidot} is equivalent to \er{lagmul4} and thus is satisfied identically. \Er{chidot} is equivalent to \ern{albetaeq}
and \ern{sigmaeq}. Hence the above equations are the same as \ern{equa} but in Hamiltonian presentation.

\section {Conclusion}

It is shown that non-barotropic magnetohydrodynamics can be derived from a variational principle of five functions.
The formalism is given in both a Lagrangian and a Hamiltonian presentation.

Possible applications include stability analysis of stationary magnetohydrodynamic configurations and its possible utilization
for developing efficient numerical schemes for integrating the magnetohydrodynamic equations.
It may be more efficient to incorporate the developed formalism in the frame work of an existing
code instead of developing a new code from scratch. Possible existing codes are described
in \cite{Mignone,Miyoshi,Narayan,Shapiro,Reisenegger}.
I anticipate applications of this study both to linear and non-linear stability analysis of known
 barotropic magnetohydrodynamic configurations \cite{VMI,Kruskal,AHH}.
 I suspect that for achieving this we will need to add additional
constants of motion constraints to the action as was done by \cite{Arnold1,Arnold2}
see also \cite{Katz,YahalomKatz,YahalomMonth}. As for designing efficient numerical schemes for integrating
the equations of fluid dynamics and magnetohydrodynamics one may follow the
approach described in \cite{Yahalom,YahalomPinhasi,YahPinhasKop,OphirYahPinhasKop}.

Another possible application of the variational method is in deducing new analytic
solutions for the magnetohydrodynamic equations. Although the equations are notoriously
difficult to solve being both partial differential equations and nonlinear, possible
solutions can be found in terms of variational variables. An example
for this approach is the self gravitating torus described in \cite{Yah3}.

One can use continuous symmetries which appear in the variational Lagrangian to derive
through Noether theorem new conservation laws. An example for such derivation which
still lacks physical interpretation can be found in \cite{Yah5}. It may be that the
Lagrangian derived in \cite{Yah} has a larger symmetry group. And of course one anticipates
a different symmetry structure for the non-barotropic case.

Topological invariants have always been informative, and there are such invariants in MHD flows. For
example the two helicities  have long been useful in research into
the problem of hydrogen fusion, and in various astrophysical scenarios. In previous
works \cite{YaLy,Yah2,Yahalomhel} connections between helicities with symmetries of the barotropic fluid equations were made. The variables of
the current variational principles may help us to identify and characterize as yet unknown topological
invariant in MHD.

\end {document}